\documentclass[prd,showpacs,showkeys,superscriptaddress,amsmath,amssymb,preprintnumbers,nofootinbib]{revtex4}

\usepackage{graphicx}       
\usepackage{dcolumn}        
\usepackage{bm}             
\usepackage{amsfonts}       

\def \be{\begin{equation}}
\def \ee{\end{equation}}
\def \bea{\begin{eqnarray}}
\def \eea{\end{eqnarray}}
\def \met{\mbox{g}}

\newcommand{\BH}{{\mbox{\tiny BH}}}
\newcommand{\NS}{{\mbox{\tiny NS}}}

\newcommand{\AH}{{\mbox{\tiny AH}}}
\newcommand{\ADM}{{\mbox{\tiny ADM}}}
\newcommand{\TTotal}{{\mbox{\tiny total}}}
\newcommand{\Self}{{\mbox{\tiny self}}}

\newcommand{\CARTESIAN}{{\mbox{\tiny C}}}

\newcommand{\diag}{\mbox{diag}}
\newcommand{\dt}{\hat{\partial}_t\,}
\newcommand{\tg}{\tilde\gamma}
\newcommand{\tA}{\tilde A}
\newcommand{\tG}{\tilde\Gamma}
\newcommand{\ttov}{\bar T}
\newcommand{\rtov}{\bar R}

\begin{document}

\def\jnl@style{\it}
\def\aaref@jnl#1{{\jnl@style#1}}

\def\aaref@jnl#1{{\jnl@style#1}}

\def\aj{\aaref@jnl{AJ}}                   
\def\apj{\aaref@jnl{ApJ}}                 
\def\apjl{\aaref@jnl{ApJ}}                
\def\apjs{\aaref@jnl{ApJS}}               
\def\apss{\aaref@jnl{Ap\&SS}}             
\def\aap{\aaref@jnl{A\&A}}                
\def\aapr{\aaref@jnl{A\&A~Rev.}}          
\def\aaps{\aaref@jnl{A\&AS}}              
\def\mnras{\aaref@jnl{MNRAS}}             
\def\prd{\aaref@jnl{Phys.~Rev.~D}}        
\def\prl{\aaref@jnl{Phys.~Rev.~Lett.}}    
\def\qjras{\aaref@jnl{QJRAS}}             
\def\skytel{\aaref@jnl{S\&T}}             
\def\ssr{\aaref@jnl{Space~Sci.~Rev.}}     
\def\zap{\aaref@jnl{ZAp}}                 
\def\nat{\aaref@jnl{Nature}}              
\def\aplett{\aaref@jnl{Astrophys.~Lett.}} 
\def\apspr{\aaref@jnl{Astrophys.~Space~Phys.~Res.}} 
\def\physrep{\aaref@jnl{Phys.~Rep.}}      
\def\physscr{\aaref@jnl{Phys.~Scr}}       

\let\astap=\aap
\let\apjlett=\apjl
\let\apjsupp=\apjs
\let\applopt=\ao

\title[Hydro-without-Hydro Framework for Simulations of Black Hole--Neutron Star Binaries]
{Hydro-without-Hydro Framework for Simulations of\\ Black Hole--Neutron Star Binaries}


\author{Carlos F. Sopuerta}
\affiliation{Center for Gravitational Wave Physics,
Penn State University, University Park, PA 16802}

\author{Ulrich Sperhake}
\affiliation{Center for Gravitational Wave Physics,
Penn State University, University Park, PA 16802}
\affiliation{Theoretisch-Physikalisches Institut, Friedrich-Schiller-Universit\"at Jena,
Max-Wien-Platz 1, 07743 Jena, Germany}

\author{Pablo Laguna}
\altaffiliation[Also at ]{Institute for Gravitational Physics and Geometry,
Departments of Astronomy \& Astrophysics and Physics}
\affiliation{Center for Gravitational Wave Physics,
Penn State University, University Park, PA 16802}

\date{\today}

\begin{abstract}
We introduce a computational framework which avoids solving explicitly hydrodynamic equations
and is suitable to study the pre-merger evolution of 
black hole -- neutron star binary systems.
The essence of the method consists of constructing a neutron star model with a black hole
companion and freezing the internal degrees of freedom of the neutron star during
the course of the evolution of the space-time geometry.
We present the main ingredients of the framework, from the formulation of the problem to the
appropriate computational techniques to study these binary systems.
In addition, we present numerical results of the construction of initial data sets and
evolutions that demonstrate the feasibility of this approach.
\end{abstract}



\preprint{CGWP/xx}

\maketitle

\section{Introduction}
The coalescence of compact binaries are among the most
important sources of gravitational waves to be detected by ground-based
laser interferometers like LIGO, TAMA, GEO, and VIRGO.  These observational efforts
have served as the primary motivation
to simulate compact binary systems. The simulations will provide
crucial information in support of the data analysis.
Most of the focus has been on the simulation of black hole (BH) binaries and neutron star
(NS) binaries, leaving aside the
BH-NS binary system (see~\cite{Taniguchi:2005fr,Faber:2005yg,Faber:2006qc} for recent progress).
 From the point of view of Numerical Relativity,
the system BH-NS exhibits a dual challenge.
It has the difficulties of evolving the geometry in the vicinity of
BHs together with the difficulties inherent to magneto-hydrodynamical
calculations.

From an astrophysical point of view, NS-BH binaries have the added interest 
of their potential relevance to gamma ray bursts.
Observations of short gamma ray bursts
(see~\cite{Gehrels:2005sg,Villasenor:2005gr,Fox:2005gr,Hjorth:2005gr}
and more recently~\cite{Berger:2005gr,Tanvir:2005gr,Barthelmy:2005gr})
suggest that the coalescence of NS binaries and/or
BH-NS binaries are the underlying mechanism in the central engine for
those bursts that last about $0.3\,s\,$.

Due to the high computational cost of simulations of BH-NS binary systems (and the
high cost of developing the appropriate computational infrastructure) it is desirable
to have a framework in which to study certain dynamical regimes of this system,
relevant to gravitational-wave physics, that admit the use of approximations
and thus facilitates a drastic reduction of the computational resources required.
In this sense, we are particularly interested in the stages of the evolution where the
predictions of post-Newtonian methods become unreliable and where numerical relativistic
simulations should take over.   The {\em hydro-without-hydro} approximation that we propose in
this paper aims at covering such part of the dynamical regime where the dynamical
timescales related with deformations of the NS due to tidal effects are much
bigger than the orbital timescales.  In this regime we can freeze most of the
hydrodynamical degrees of freedom and evolve a finite number of them by using
appropriate approximations, focusing the attention on the radiation reaction
effects in the orbit.  It is this reduction of degrees of freedom which avoids the use
of hydrodynamical computations.
Such an approach may also be relevant for extreme-mass-ratio
binary systems whose dynamics are driven by radiation-reaction effects.
The main goal of this paper is then to describe the setup for this type of
simulations and test, by extending an existing code for BH evolutions,
the basic ingredients of this framework.  These ingredients can be summarized as follows:

\paragraph{Spacetime description:} We use a full numerical
relativity description of the spacetime based on the 3+1 BSSN
formulation (Baumgarte, Shapiro~\cite{Baumgarte:1998te}, and Shibata,
Nakamura~\cite{Shibata:1995we}) of Einstein's equations as implemented
in the 3D numerical relativity code MAYA (described in detail
in~\cite{Shoemaker2003,Sperhake2004}).  The MAYA code is written
by using the computational toolkit CACTUS~\cite{cactus} and the Mesh
Refinement package CARPET~\cite{carpet,Schnetter:2004sh}.  The
key extension of this code regarding this project is the introduction
of the matter source terms associated with the NS and the
associated infrastructure to evolve the matter degrees of freedom.

\paragraph{Construction of Initial Data:}
Initial data is constructed by
superposing a Schwarzschild BH in Kerr-Schild (KS)
coordinates with a boosted Tolman-Oppenheimer-Volkoff (TOV)
\cite{Tolman1939, Oppenheimer1939} model in
a coordinate system that has similarities with the ingoing-Eddington-Finkelstein
coordinate system.  The TOV model describes the equilibrium configuration of
a stationary, relativistic star in spherical symmetry and thus represents
a convenient description of the matter sources in the BH-NS system at hand.
The superposition is carried out by using a procedure analogous to that used
for the construction of binary BH initial data initial data in
KS coordinates~\cite{Matzner:1998pt}.  Even though the resulting
initial data do not satisfy the constraint equations exactly,
the constraint violations are comparatively small because the individual
BH and NS models do represent exact solutions of the Einstein field equations.
In this work we will investigate in detail the properties of these initial data
from a physical and mathematical point of view. We will also discuss potential
modifications of the data with regard to their repercussions on the
physical content.  Finally we illustrate how adequate evolutions, even
in the case of more extreme mass ratios, are facilitated by the modern
infrastructure of numerical relativity.

\paragraph{Description and motion of the NS:} In the {\em hydro without
hydro} approximation pursued in this work, we need to address
two issues relating to the description of a particular type of energy-momentum
distribution: the update of the matter sources
that appear in the BSSN evolution equations and the motion of such a
distribution.  In a full numerical relativity setup this is not necessary
as the complete evolution of the matter sources is taken care of by the
Einstein field- and the energy-momentum conservation equations.
In reducing the degrees of freedom in the way we are proposing, one is
usually left with an approximation scheme in which the {\em internal}
and {\em external} motions are cleanly separated, even though
they are coupled in general. The {\em external} motion consists
in the motion of a reference point of the matter distribution, typically
a relativistic generalization of the Newtonian concept of center of
mass, whereas the {\em internal} motion consists in the evolution of
the parameters describing, for instance, the deformations of the matter
distributions due to tidal deformations.
As a consequence of reducing the number of degrees of freedom, the
description of the motion of the energy-momentum distribution
in our setup consists of a set of ordinary differential equations (ODEs).
The matter sources, as they appear in the evolution equations for
the spacetime geometry are thus updated by merely substituting
the values of the energy-momentum quantities after the motion of the
matter has been determined.  In this paper, we consider the simplest
such description of the NS, namely that of a star {\em rigidly} moving along a
prescribed trajectory.  By using this simplification
we ignore internal motions as well as radiation-reaction
and finite size effects due to the {\em external} gravitational field.
More specifically, the matter sources are prescribed by taking
particular density and pressure profiles from a TOV model which are
going to be maintained along the evolution.   The only way in which the
sources change is through the {\em rigid} bulk motion of the center of
the TOV model along a fixed trajectory. The reason for considering such
a simplified model is to perform
a feasibility study of this method and
thus lay the foundation for future studies of more realistic
physical configurations of the BH-NS binary system.
A similar approach to that proposed here has recently
been used by Bishop {\em et al}~\cite{Bishop:2003bs}
in the framework of a characteristic
formulation of the Einstein equations.

The plan of this paper is as follows: In section~\ref{sec:theory}
we present the mathematical description of the different
ingredients of our framework: spacetime geometry and dynamics
(\ref{spacetime}), the descriptions of the BH and the NS
(\ref{bhdescription} and~\ref{sec:tov} respectively), the
initial data for the BH-NS system (\ref{initialdata}), the
dynamics of the NS (\ref{nsmotion}).
In section~\ref{numericalframework}
we describe the computational framework and report on
the results of numerical tests we have performed to analyze
the initial data (subsection~\ref{analysisinitialdata}) and to
test time evolutions with matter sources (\ref{timeevolutions}).
We conclude in section~\ref{sec:conclusions} with a discussion
of our findings. For the purpose of presenting the relativistic
equations we set $c = G = 1$ throughout this work.

\section{Theoretical Foundations \label{sec:theory}}
In this section we describe the different ingredients that constitute the
mathematical framework that we are proposing for the numerical description
in the time domain of  BH-NS binary systems.
For this purpose we need to consider both, the generation of initial data,
and the evolution of these data. In the construction of initial data
we closely follow a procedure used
for a binary BH systems: the superposition of two BH
solutions in KS form (cf.\, \cite{Matzner:1998pt,
Marronetti:2000rw,Bonning:2003im,Sperhake:2005uf}).
This technique is based on the properties of the KS form of a single BH
metric and its invariance under Lorentz transformations of the coordinates.
Then, starting from two single BHs in KS coordinates (and different coordinate
origin) a Lorentz boost is applied to each of them and then, by identifying
the coordinate systems, a superposed metric is constructed.
This metric is no longer an exact solution of the Einstein field equations
(the constraint are not satisfied), but it is an approximation which improves as
the BH separation increases.    This type of data has been shown
to work well in numerical relativity simulations~(see, e.g.~\cite{Sperhake:2005uf}).
It has the advantage that one can consider Kerr BH initial data without the need to
worry about spurious radiation that may originate in initial data constructions based
on conformally-flat slicings.

The evolution of the data is performed in analogy to that of BH-data, with
one exception: the addition of matter source terms on the right hand side
of the evolution equations. We thus obtain a fully non-linear evolution
of the geometry of the spacetime. In contrast, we substantially reduce the
degrees of freedom of the matter sources. We will discuss in detail in this
section, how this enables us to obtain an approximate evolution of the
matter data.

\subsection{Description of the Spacetime Geometry}\label{spacetime}
We begin our description of the mathematical model with
the spacetime geometry.  In this work we follow the
standard 3+1 splitting of Einstein's equations based on the
Arnowitt-Deser-Misner (ADM) formulation~\cite{adm:1962ok}
(see~\cite{Wald:1984rw,York:1979jw} for details).
There it is assumed that the spacetime has topology
$\mathbb{R}\times\Sigma$, and hence can be foliated by the level
surfaces $\Sigma(t)$ of a  function $t(x^\mu)$.  The unit normal to
the hypersurfaces $\Sigma(t)$, $n^\mu=-\alpha\nabla^\mu t$, satisfies
$\met_{\mu\nu}n^{\mu}n^{\nu}=-1\,$. It can be used to decompose the
four-metric according to
\begin{equation}
\met_{\mu\nu} = h_{\mu\nu}  - n_{\mu}n_{\nu} \,,
\end{equation}
where $h^\mu{}_\nu$ is the projector orthogonal to $n^\mu$ ($h^\mu{}_\nu
n^\nu=0$).  A vector field $t^{\mu}$ threading the spacetime is
introduced, $t^{\mu}\nabla_{\mu}t=1\,,$ in terms of which we can
introduce the lapse function and shift vector through the relations
$\alpha = - t^{\mu} n_{\mu}$ and $\beta^{\mu} = t^{\mu} - \alpha
n^{\mu}$ respectively.  Then, using a system of spatial coordinates
$\{x^i\}$ adapted to the hypersurfaces $\Sigma(t)$, the metric takes
the well-known form:
\begin{equation}
  ds^2 = -\alpha^2dt^2+\gamma_{ij}(dx^i+\beta^idt)(dx^j+\beta^jdt) \,,
         \label{2m1}
\end{equation}
where $\gamma_{ij}=h_{ij}$ are the components of the spatial metric of the
hypersurfaces $\Sigma(t)$.   The extrinsic curvature of $\Sigma(t)$
is defined as
\begin{equation}
  K_{\mu\nu} = -\textstyle{\frac{1}{2}}{\pounds}_n h_{\mu\nu} \,, \label{kmunu}
\end{equation}
where $\pounds$ denotes the Lie derivative operator. We thus obtain
\begin{equation}
  \hat{\partial}_t \gamma_{ij} = -2\alpha K_{ij} \,,
\end{equation}
where we have introduced the time derivative operator:
\begin{equation}
  \hat{\partial}_t = \partial_t-{\pounds}_\beta\,. \label{that}
\end{equation}
%

Because we are dealing with a physical system that contains matter, we
also need to apply the 3+1 splitting to the
energy-momentum tensor $T^{\mu\nu}$.   This can be done
by decomposing
$T^{\mu\nu}$ with respect to the unit normal $n_\mu$:
\begin{equation}
  T^{\mu\nu} = E n^\mu n^\nu + P h^{\mu\nu} + 2 J^{(\mu}n^{\nu)}+
  \Pi^{\mu\nu}\,,  \label{emt}
\end{equation}
where the different quantities that appear in this expression are defined
as follows:
\begin{eqnarray}
  E & = & T^{\mu\nu}n_\mu n_\nu\,, \label{Energy} \\
  P & = & \textstyle{\frac{1}{3}}h_{\mu\nu}T^{\mu\nu}\,,\label{Pressure} \\
  J^\mu   & = & -h^\mu{}_\nu T^{\nu\tau}n_\tau\,, \label{Current} \\
  \Pi_{\mu\nu} & = & h_{\mu\tau} h_{\nu\sigma} T^{\tau\sigma}
        - P h_{\mu\nu}\,. \label{Stresses}
\end{eqnarray}

Applying the 3+1 splitting to Einstein's equations we obtain
the {\em Evolution equations}:
\begin{eqnarray}
  \dt \gamma_{ij} &=& - 2 \alpha K_{ij}\,, \label{dgdt} \\
  \dt K_{ij} &=& - D_i D_j \alpha + \alpha \left[R_{ij} + K K_{ij}
      - 2 K_{ik} {K^k}{}_j
      -\Pi_{ij}-\frac{1}{2}(E-P)\gamma_{ij}\right]\,, \label{dKdt}
\end{eqnarray}
and the {\em Constraint equations}:
\begin{eqnarray}
  {\cal H} & \equiv & R + K^2 - K_{ij} K^{ij} - 2E = 0\,,
       \label{Hconstraint}\\
  {\cal P}^i & \equiv & D_j (K^{ij} - \gamma^{ij} K) - J^i = 0\,.
       \label{Pconstraint}
\end{eqnarray}
Here, $D_i$ is the covariant derivative associated with the 3-metric
$\gamma_{ij}$, $R_{ij}$ is the three-dimensional Ricci tensor,
$R=\gamma^{ij}R_{ij}$ the Ricci scalar, and $K$ is the trace of
$K_{ij}$.  Equations (\ref{dgdt},\ref{dKdt}) and
(\ref{Hconstraint},\ref{Pconstraint}) constitute the basic
3+1 setup and are known as the ADM equations~\cite{adm:1962ok}.
We are going to use a modification of this formulation due to
Shibata and Nakamura~\cite{Shibata:1995we}, and Baumgarte and
Shapiro~\cite{Baumgarte:1998te}, now known as the BSSN 3+1 formulation of
Einstein's equations, which has been found to result in vastly improved
stability properties in numerical simulations compared with the ADM equations.
The way in which the BSSN formulation is introduced from the ADM formulation
is as follows:
We introduce new variables based on a trace decomposition of the
extrinsic curvature and a conformal rescaling of both the metric and
the extrinsic curvature. The trace-free part $A_{ij}$ of the extrinsic
curvature is defined by
\begin{equation}
  A_{ij} = K_{ij} - \frac{1}{3} \gamma_{ij} K\,.  \label{defA}
\end{equation}
Next, one introduces a conformal metric $\tg_{ij}$ in terms of the
physical metric by
\begin{equation}
  \gamma_{ij} = \psi^4 \tg_{ij}\,.
\end{equation}
The value of the conformal factor $\psi$ can be fixed by
demanding the determinant of the conformal metric $\tg_{ij}$ to be unity:
\begin{equation}
  \psi = \gamma^{1/12}\,,~~~~~
  \tg_{ij} = \psi^{-4}\gamma_{ij} = \gamma^{-{1/3}}\gamma_{ij}\,,~~~~~
  \tilde\gamma = 1\,,
\end{equation}
where $\gamma$ and $\tilde\gamma$ are the determinants of $\gamma_{ij}$ and
$\tg_{ij}$ respectively.   Then, instead of $\gamma_{ij}$ and $K_{ij}$
we can use the variables
\begin{eqnarray}
  \phi &=& \ln \psi = \frac{1}{12} \ln \gamma\,,  \\
  K &=& \gamma_{ij} K^{ij}\,, \\
  \tg_{ij} &=& e^{-4\phi} \gamma_{ij}\,, \\
  \tA_{ij} &=& e^{-4\phi} A_{ij}\,,
\end{eqnarray}
where $\tg_{ij}$ has determinant $1$ and $\tA_{ij}$ has vanishing trace.
Moreover, we also introduce the following conformal connection functions
\begin{equation}
  \tG^i = \tg^{jk} \tG^i{}_{jk} = - \partial_j \tg^{ij}\,, \label{defG}
\end{equation}
where $\tG^i{}_{jk}$ denotes the Christoffel symbols associated with the
conformal metric.  To obtain the second equality we have to use the fact
that the determinant of the conformal spatial metric, $\tilde \gamma$,
is unity.

The variables used in the BSSN formulation are: $\phi$, $K$, $\tg_{ij}$,
$\tA_{ij}$, and $\tG^i$.  The set of evolution equations for these
variables that we are going to use, and which can be derived from the
previous relations, is given by
\begin{eqnarray}
  \dt \tg_{ij} & = & -2\alpha \tA_{ij}\,,  \label{dtgdt} \\
  \dt \phi     & = & -\frac{1}{6} \alpha K\,, \label{dphidt} \\
  \dt \tA_{ij} & = & e^{-4\phi} \left[ - \left(D_i D_j \alpha\right)^{TF}
               + \alpha\left( R_{ij}^{TF}
               -   8\pi\Pi_{ij}\right)\right]
               +   \;\alpha (K \tA_{ij} - 2 \tA_{ik} \tA^k{}_j)\,,
               \label{dtAdt} \\
  \dt K & = & - D^i D_i \alpha + \;\alpha \left[\tA_{ij} \tA^{ij}
               + \frac{1}{3} K^2
               + 4\pi(E+3P) \right]\,, \label{dtrKdt} \\
  \partial_t \tG^i & = & \beta^j \partial_j \tG^i - \tG^j \partial_j \beta^i
               -    2\tA^{ij}\partial_j \alpha + 2\alpha\left(
               \tG^i_{jk}\tA^{jk}
               +    6\tA^{ij}\partial_j\phi - \frac{2}{3}\tg^{ij}\partial_j K
               -    8\pi e^{4\phi} J^i\right) \\ \nonumber
               &  + & \tg^{jk}\partial_j \partial_k\beta^i
               + \frac{1}{3}\tg^{ij}\partial_j\partial_k \beta^k
               +   \frac{2}{3} \tG^i \partial_j \beta^j + \left(\chi
               + \frac{2}{3}\right)
               \left( \tG^i - \tg^{jk}\tG^i_{jk} \right) \partial_j \beta^j\,.
               \label{dtGdt}
\end{eqnarray}
Here the superscript $TF$ denotes the trace-free part (with
respect to the metric $\gamma_{ij}$).
In this context, it is important to note that $\phi$ is defined in terms of the
determinant of the 3-metric $\gamma_{ij}$, which is a scalar density of weight
$2$, and that $\tg_{ij}$ and $\tA_{ij}$ are tensor densities of weight $-2/3\,.$
Finally, $\chi$ is a free parameter [it multiplies a quantity that
vanishes at the analytic level by virtue of the definition of the
quantities $\tG^i$~(\ref{defG})] that has been set to $2/3$ for all
simulations reported here.  This value has been suggested by numerical experiments.

\subsection{BH description: Kerr-Schild form of a Schwarzschild BH}
\label{bhdescription}

Although spinning BHs are likely to be more interesting from an
astrophysical point of view, especially in the case of extreme-mass-ratio binaries,
in this work we restrict our attention to non-rotating Schwarzschild BHs.
We do note, however, that for the case of KS coordinates,
the fundamental properties of spinning and non-spinning black holes
are the same from a numerical point of view (see, for example, the successful
evolutions of single Kerr BHs obtained in~\cite{Yo2002}). This feature
represents a key motivation in our choice of coordinates and is
not obviously satisfied by coordinates which are not adapted to the spinning cases
(for instance, all the coordinate systems that assume an initially conformally-flat
spatial geometry).

In this section we describe the contribution to the initial data
of the BH, that is we need to prescribe initial data for the functions
$(\phi,K,\tg_{ij},\tA_{ij},\tG^i)\,.$ These arise from the KS form of the
metric, which can be conveniently written as the sum of a flat space metric
plus a term that factorizes as the product of a light-like vector with
itself according to
\begin{equation}
  ds^2 = (\eta_{\mu\nu}+2H\ell_\mu\ell_\nu)dx^\mu dx^\nu\,. \label{bh1}
\end{equation}
Here $\eta_{\mu\nu} = \diag(-1,1,1,1)$ is the Minkowski metric in
Cartesian coordinates $\{t,x^i\}$ and
$\ell^\mu$ is a null vector field
\begin{eqnarray}
  \ell_\mu dx^\mu &=& -dt - \frac{x_i}{r}dx^i \,,~~~r^2
         = \delta_{ij}x^ix^j\,, \label{bh2}
\end{eqnarray}
where $x_i=x^i$.  This light-like vector field has been chosen so that it
corresponds to ingoing light rays.
Finally, $H$ is a scalar given by
\begin{equation}
  H = \frac{M_\BH}{r}\,, \label{bh3}
\end{equation}
where $M_\BH$ is the BH mass.

The metric quantities are then independent of the time $t$ which
slices the spacetime
\begin{equation}
  \alpha^2_\BH = \frac{1}{1+2H}\,,~~~~~
  \beta^i_\BH = \frac{2H}{1+2H}\ell^i\,, \label{ks1}
\end{equation}
\begin{equation}
\gamma_{ij}^\BH = \delta_{ij} + 2H\ell_i\ell_j\,,~~~~~
\gamma^{ij}_\BH = \delta^{ij} - \frac{2H}{1+2H}\ell^i\ell^j\,,
\end{equation}
\begin{equation}
\gamma_\BH = \det(\gamma_{ij}^\BH) = 1+2H = \alpha^{-2}_\BH \,. \label{ks2}
\end{equation}
%

\subsection{NS matter description: The Tolman-Oppenheimer-Volkoff stellar model}
\label{sec:tov}

The main approximation in our description of the BH-NS binary system
lies in the description of the NS.  The main goal of this
approximation is to provide a framework in which one avoids
solving the hydrodynamical
equations governing the NS matter fields while at the same time being able to
have a reasonable approximation of the main dynamical aspects of the evolution
of such a system.  The implementation of this idea requires two ingredients:
(i)  The prescription of the profiles for the different matter variables and
the construction of initial data for the geometry. (ii) The evolution of the
matter profiles.  Regarding the second point, in this paper we take the
simplest description, namely, to consider {\em rigid} profiles that follow
a prescribed trajectory.  In subsection~\ref{nsmotion} we briefly discuss how
this description can be refined in future work to obtain physically more
realistic models.

With regard to the first point, the matter profiles for the NS matter
variables are obtained by using an exact solution representing an isolated
NS star, more precisely a TOV model.  With regard to the eventual
superposition with the BH, we will apply two modifications to the
standard TOV metric: i) a transformation to KS-like coordinates,
and ii) apply a boost transformation to account for the motion of the star
relative to the black hole.
Our starting point is the TOV metric in Schwarzschild coordinates
\begin{equation}
  \met_{\alpha\beta}dx^\alpha dx^\beta = -A^2(\rtov)d\ttov^2 + B^2(\rtov)
        d\rtov^2+ \rtov^2 d\Omega^2 \,, \label{eq: TOV_metric}
\end{equation}
where $d\Omega^2 = d\theta^2+\sin^2\theta d\varphi^2\,.$
The energy momentum tensor of a TOV star is that of a perfect fluid
and is given by
\begin{equation}
  T_{\alpha\beta} = (\rho+p)\, u_\alpha u_\beta + p\,\met_{\alpha\beta}\,,
  \label{tmunu}
\end{equation}
where $u_\alpha=\left(-A,0,0,0\right)$ is the 4-velocity of the
fluid elements. It is convenient to introduce the {\em mass}
function $m(\rtov)$ by
\begin{equation}
  B^{-2} = 1-2m/\rtov\,.
\end{equation}
Then, the TOV equations are:
\begin{eqnarray}
& & \frac{A'}{A}=\frac{m+4\pi p \, \rtov^3}{\rtov\,
      (\rtov-2m)} = -\frac{p'}{\rho+p}\,, \label{toveq1}  \\
& & m' = 4\pi\rho \, \rtov^2 \,, \label{toveq2}
\end{eqnarray}
where a prime denotes differentiation with respect to $\rtov\,.$
We are going to consider the case of a barotropic Equation of
State (EoS), specifically a polytropic one:
\begin{equation}
p= k\rho^\Gamma\,,~~~\Gamma = 1 + \frac{1}{n},   \label{toveq3}
\end{equation}
where the polytropic coefficient $k$ and the polytropic index $n$ are constants.
Equations (\ref{toveq1},\ref{toveq2}) determine a one-parameter
family of TOV solutions, which can be parametrized by the value
of the central density $\rho^{}_{\rm c} = \rho(\rtov=0)\,.$  The sound speed
of the matter distribution is given by $c_s^2= dp/ d\rho\,.$

Bearing in mind the eventual superposition of the BH and NS metrics in
analogy to the construction of KS binary BH data, we are going to apply
coordinate transformations to the TOV solution that resemble the coordinate
transformations used in the BH case.   In the case of BHs, the first coordinate
transformation aims at changing the time coordinate so that it is adapted
to light-like geodesics, typically ingoing ones.  For a Schwarzschild BH,
this corresponds to the KS coordinate system.  In the case of a NS we follow
the same idea.  A straightforward calculation shows that after applying
the coordinate transformation
\begin{eqnarray}
  d\hat{T}= d\ttov + (B/A-1)d\rtov\,,~~~
  dR = d\rtov\,. \label{schtoief}
\end{eqnarray}
the ingoing radial light-like geodesics are given by $\hat{T}+R\,$.
This transformation, however, is not suitable for a three-dimensional
numerical treatment because it leads to a coordinate singularity at the centre
of the star when we transform to Cartesian coordinates [by using the
transformation given below in equation~(\ref{ieftocartesian})].
It is possible, though, to find a coordinate transformation
analogous to (\ref{schtoief}) that satisfies all regularity requirements
at the stellar centre.  For this purpose we define the new coordinate
time $T$ by
\begin{equation}
  dT = d\ttov + \left(\frac{B}{A}-\frac{1}{AB}\right)d\rtov\,.~~~
  \label{ourschtoief}
\end{equation}
The resulting line element is then given by:
\begin{equation}
ds^2 = -A^2 dT^2 + 2AB(1-B^{-2})dT dR + (2-B^{-2})dR^2 + R^2d\Omega^2\,.
\label{tovief}
\end{equation}
By virtue of Birkhoff's theorem, the geometry exterior to the star is
necessarily described by the Schwarzschild metric.
In order to combine the interior and exterior metric, we need
to match them in a smooth way at the stellar surface.
It is a remarkable property of the interior metric (\ref{tovief})
that it still smoothly matches to the exterior (KS) solution
and thus leads to coordinates adapted to the ingoing null geodesic structure
outside the star. This follows straightforwardly from performing the
substitution $A^2=B^{-2}=1-2M_\NS/R\,,$  where $M_\NS$ is the NS mass.  In
consequence, the interior and exterior geometries match smoothly (in the sense
of the Lichnerowicz junction conditions~\cite{Lichnerowicz:55al,Lichnerowicz:1971al}),
at a hypersurface $R=R^{}_\NS=\mbox{constant}\,,$  provided the TOV model
satisfies the following conditions
\begin{eqnarray}
A^2|_{R=R^{}_\NS} = 1-2M_\NS/R^{}_\NS\,,~~~
p(R^{}_\NS)=0\label{matching}\,,
\end{eqnarray}
which have the obvious consequences: $m(R^{}_\NS)=M_\NS$ and
$B^{-2}|_{R=R^{}_\NS}=1-2M_\NS/R^{}_\NS$ (or equivalently
$AB|_{R=R^{}_\NS}=1$), and clearly $R^{}_\NS$ represents the NS surface.

Next, we apply the coordinate change from
spherical $(T,R,\theta,\varphi)$ to the Cartesian-like
coordinates $(T,X^i)\,$. These coordinates are related by
\begin{equation}
  (X^i) = (X,\,\,Y,\,\,Z) = (R\sin\theta\cos\varphi,\,\,
     R\sin\theta\sin\varphi,\,\, R\cos\theta) \,. \label{ieftocartesian}
\end{equation}
The resulting line element is
\begin{eqnarray}
  ds^2 = -A^2 dT^2 + 2AB(1-B^{-2})dT\frac{X_i}{R}dX^i +
  \left[\delta_{ij}+(1-B^{-2})\frac{X_iX_j}{R^2}\right]dX^idX^j \,,
  \label{cartesian}
\end{eqnarray}
where $R$ is considered a function of $X^i$ given by $R^2=\delta_{ij}X^iX^j\,$,
and $X_i=X^i\,$.  At this point, we can see
that $B\rightarrow 1$ as $R\rightarrow 0\,$,
i.\,e.\,the metric is regular at the origin $X^i =0\,$ of the stellar object.
This would not have been the case for the coordinate
transformation (\ref{schtoief})
which would have lead to a cusp at the origin of the star.

The ADM metric variables associated with the slicing $\{T=\mbox{constant}\}$
are:
\begin{eqnarray}
  \alpha^2_\CARTESIAN    & = & \frac{A^2B^2}{2-B^{-2}} \,, \label{alphac} \\
  \beta^i_\CARTESIAN     & = & AB\frac{1-B^{-2}}{2-B^{-2}}\frac{X^i}{R}\,,
  \label{betac} \\
  \gamma_{ij}^\CARTESIAN & = & \delta_{ij} + (1-B^{-2})\frac{X_iX_j}{R^2}\,,
  \label{3metricc}
\end{eqnarray}
where we have used the tag '$\CARTESIAN$' to distinguish the quantities associated
with this particular slicing.  We now address the next step in the construction
of the superposed geometry, the Lorentz boost.
We first note, that the NS metric (\ref{cartesian}) cannot be cast
in the KS form~(\ref{bh1}),  which is preserved
under Lorentz transformations.  While the NS metric can still be written in the form
$\eta_{\mu\nu}+H_{\mu\nu}$, the resulting $H_{\mu\nu}$ cannot be factorized in
terms of a null vector field, where $H_{\mu\nu}\rightarrow 0$ as $R\rightarrow\infty$.
On the other hand, the form $\eta_{\mu \nu} + H_{\mu \nu}$ is still Lorentz invariant.
We thus apply a coordinate change from the Cartesian-like coordinates
$(T,X^i)$ of Eq.\,(\ref{cartesian}) to a different set of Cartesian-like
coordinates $(t,x^i)$ via a Lorentz boost.  These two sets of
coordinates are related by the following expressions:
\begin{eqnarray}
  T & = & \gamma \left[t - \delta_{ij}v^i(x^j-z^j)\right] \,, \label{boost1} \\
  X^i & = & -\gamma t v^i + \left[\delta^i_j + (\gamma-1)\frac{v^iv_j}
  {v^2} \right](x^j-z^j)  \,, \label{boost2}
\end{eqnarray}
where $v^i$ are the components of the boost velocity ($v_i = v^i$),
$v^2$ is $\delta_{ij}v^iv^j\,$, $\gamma$ is $(1-v^2)^{-1/2}\,$,
and $z^i$ are the spatial components of the compact object's location.
In the course of the evolution, the position vector $z^i$ needs to be evolved
by solving the equations governing the external dynamics of the
NS.  In this paper it is going to be a prescribed trajectory.

After carrying out the transformation (\ref{boost1},\ref{boost2}) the ADM metric
variables associated with the slicing $\{t=\mbox{constant}\}$ can
be written as
\begin{eqnarray}
\alpha^2_\NS    & = & \alpha^2_\CARTESIAN\gamma^{-2}
    \left[(1-v_i\beta^i_\CARTESIAN)^2 -
    \alpha^2_\CARTESIAN\gamma^{ij}_\CARTESIAN v_iv_j \right]^{-1}
    \,, \label{alpha} \\
\beta^i_\NS     & = & -\frac{\gamma}{1+\gamma}v^i + \gamma\alpha^2_\NS
    \left[\alpha^{-2}_\CARTESIAN\left(1-v_i\beta^i_\CARTESIAN\right)
    \left(\beta^i_\CARTESIAN-\frac{\gamma}{1+\gamma}v^i\right)
    +\gamma^{ij}_\CARTESIAN v_j\right] \,, \label{beta} \\
\gamma_{ij}^\NS & = & \gamma_{ij}^\CARTESIAN -2\gamma
    v^{}_{(i}\gamma_{j)k}^\CARTESIAN
    \left(\beta^k_\CARTESIAN-\frac{\gamma}{1+\gamma}v^k\right) +
    \gamma^2\left[\gamma_{kl}^\CARTESIAN\left(\beta^k_\CARTESIAN
    -\frac{\gamma}{1+\gamma}v^k\right)
    \left(\beta^l_\CARTESIAN-\frac{\gamma}{1+\gamma}v^l\right)
    - \alpha^2_\CARTESIAN\right]
    v_iv_j \,, \label{3metric}
\end{eqnarray}
where $\alpha_\CARTESIAN\,$, $\beta^i_\CARTESIAN\,$,
and $\gamma_{ij}^\CARTESIAN$ are given in equations
(\ref{alphac})-(\ref{3metricc}) in terms of the coordinates $(T,X^i)\,$.
With expressions (\ref{alpha})-(\ref{3metric}) we complete the geometric
description of the NS.  They will be used below to construct the
initial data by superposing the BH and the NS solutions.

We denote the matter source terms associated with the foliation
$\{t = const.\}$, by $(E_\NS,P_\NS,J^i_\NS,\Pi^\NS_{ij})$.
These can be obtained
from the expressions~(\ref{Energy})-(\ref{Stresses}), bearing in mind
that the normal to this foliation is $n_\mu dx^\mu = -\alpha_\NS dt$
and the components of the fluid four-velocity are
\begin{equation}
u^t = \gamma u^T\,,~~~ u^i = u^t v^i\,, \label{fluidvelocity}
\end{equation}
where $u^T = A^{-1}\,$.  Then, the result is
\begin{eqnarray}
E_\NS    & = & (\rho+p)\left(\frac{\alpha_\NS\gamma}{A}\right)^2 - p\,, \label{ENS}  \\
P_\NS    & = & \frac{1}{3}\left[(\rho+p)\left(\frac{\alpha_\NS\gamma}{A}\right)^2
           -   \rho +2p  \right]\,, \label{PNS} \\
J^i_\NS  & = & \alpha_\NS\frac{\gamma^2}{A^2}(\rho+p)(\beta^i_\NS+v^i)\,, \label{JNS} \\
\Pi^\NS_{ij} & = & (\rho+p)\frac{\gamma^2}{A^2}\left(\gamma_{ik}^\NS\gamma_{jl}^\NS
         - \textstyle{\frac{1}{3}}\gamma_{ij}^\NS\gamma_{kl}^\NS\right)
(\beta^k_\NS+v^k)(\beta^l_\NS+v^l)\,,   \label{PiNS}
\end{eqnarray}
where $\alpha_\NS\,$, $\beta^i_\NS\,$, and $\gamma_{ij}^\NS$ are given by
(\ref{alpha})-(\ref{3metric}), and $\rho$ and $p$ are the TOV
energy density and isotropic pressure.

\subsection{Superposition of the BH and NS: Initial Data and BSSN Source Terms}
\label{initialdata}

We now have all the necessary ingredients to superpose the BH and NS spacetimes
given above.  Because of the presence of matter terms, this process involves two steps:
(i) the construction of the geometric quantities associated with the
BH-NS system
and, (ii) the construction of the matter source terms associated with that
geometry.

First, we superpose the geometries in complete analogy with the
BH binary case (cf.\,\cite{Sperhake:2005uf,Kelly:2004bk}).
Let us consider, for this purpose, the BH metric given by
expressions (\ref{bh1})-(\ref{bh3}) and the NS metric given
by expressions (\ref{alpha})-(\ref{3metric}).  By identifying the
two spacetimes' coordinate systems, the spatial metric and extrinsic
curvature of the superposition are given by
\begin{eqnarray}
  \gamma^{}_{ij} & = & \gamma^\BH_{ij} + \gamma^\NS_{ij} -
      \delta^{}_{ij}\,,
      \label{3metricsup} \\
  K^i{}_j & = & K^i_{\!\BH\,j} + K^i_{\!\NS\,j}\,. \label{extcursup}
\end{eqnarray}
Moreover, the lapse and shift associated with the superposition are
given by the following expressions:
\begin{eqnarray}
  \alpha & = & ( \alpha^{-2}_\BH + \alpha^{-2}_\NS - 1 )^{-1/2}\,,
         \label{lapsesup}\\
  \beta^i & = & \gamma^{ij}\left( \beta_j^\BH + \beta_j^\NS\right)\,.
         \label{shiftsup}
\end{eqnarray}
Second, we need to obtain from the NS description the
expressions for the matter sources that appear on the right-hand sides
of the BSSN equations (\ref{dtgdt})-(\ref{dtGdt}).  To that end, we
consider the following form of the energy-momentum tensor
\begin{equation}
  T^{\mu\nu} = (\rho+p)U^\mu U^\nu + p \met^{\mu\nu}\,,
\end{equation}
where $\met^{\mu\nu}$ is the inverse of the global metric given in
equations~(\ref{3metricsup})-(\ref{shiftsup}), $\rho$ and $p$ are the TOV energy
density and isotropic pressure written in the coordinates $(t,x^i)$ of
equations~(\ref{boost1},\ref{boost2}), and $U^\mu = U^t(1,v^i)$
is the fluid velocity.  The time component $U^t$ of the fluid velocity
is determined in terms of the boost velocity $v^i$ and the global geometry
by imposing the normalization condition
$g_{\alpha \beta}U^{\alpha}U^{\beta}=-1$.
The resulting expression is
\begin{equation}
  \frac{1}{(U^t)^2} = \alpha^2 - \gamma_{kl}(\beta^k+v^k)(\beta^l+v^l).
\end{equation}
Following the same procedure as above, we can compute the matter
quantities $(E,P,J^i,\Pi_{ij})$ associated with the NS in the superposed BH-NS
spacetime. We note that the normal to the foliation is now $n_\mu dx^\mu = -\alpha dt$.
We thus obtain:
\begin{eqnarray}
  E   & = & (\rho+p)\left(\alpha U^t\right)^2 - p\,, \\
  P   & = & p + \frac{1}{3}(\rho+p)\left(U^t\right)^2\gamma_{ij}(\beta^i+v^i)
            (\beta^j+v^j) = p -\frac{1}{3}(E-\rho) \,, \\
  J^i & = & (\rho+p)\alpha\left(U^t\right)^2(\beta^i+v^i)\,, \\
  \Pi_{ij} & = & (\rho+p)\left(U^t\right)^2\left(\gamma_{ik}\gamma_{jl}
         - \textstyle{\frac{1}{3}}\gamma_{ij}\gamma_{kl}\right)
         (\beta^k+v^k)(\beta^l+v^l)\,,
\end{eqnarray}
where $\alpha\,$, $\beta^i\,$, and $\gamma_{ij}$ are given by
(\ref{lapsesup}), (\ref{shiftsup}), and (\ref{3metricsup}) respectively,
and $\rho$ and $p$ are the TOV energy density and isotropic pressure.
These expressions are very close in form to the expressions we obtained for
a single NS, the main difference being that the geometric objects now refer
to the superposed spacetime instead of the TOV spacetime.

It is clear that these initial data only satisfy the constraints in
the limit of infinite BH-NS separation.
The constraint violations inherent to this construction, however,
are expected to be small. If, for example, one considers this
type of superposed data as the background metric fields for
the conformal transverse-traceless method of Lichnerowicz, York and
others~\cite{Lichnerowicz:1944al,York:1971hw,
York:1972sj,York:1973jw,Cook:2000vr},
the conformal factor obtained from solving the Hamiltonian constraint
is close to unity throughout the whole spacetime~\cite{Bonning:2003im}.
Furthermore the quality of the data can always be enhanced by increasing
the distance between the BHs.  Finally, this data is believed to
avoid spurious radiation contamination, and in contrast to conformally-flat
data, it is possible to introduce Kerr BH solutions in a straightforward way.

Bonning {\em et al}~\cite{Bonning:2003im} have further shown
that the superposed binary BH initial data lead to the correct (Newtonian)
binding energy in the Newtonian limit.  We can ask the same question about the
prescription we have presented for BH-NS binaries.  The answer is negative, and,
in essence, the reason is that the way in which the NS has been introduced does not
account for the deformations caused by the BH gravitational field.
In the case of BH binaries the binding energy is defined as
$E_b = E^{\TTotal}_{\ADM} - E^{1}_\BH - E^2_\BH$,
where the total ADM energy is $M^1_\BH+M^2_\BH$ and the individual
energies are defined in terms of the apparent horizon masses: $E^I_\BH
= M^I_\AH$ ($I=1,2$). In the Newtonian limit ($GM/(c^2\ell)\ll
1\,,$ where $\ell$ denotes the coordinate separation) they are given
by~\cite{Bonning:2003im}: $E^1_\BH \approx M^1_\BH[1+M^2_\BH/(2\ell)]$ and
$E^2_\BH \approx M^2_\BH[1+M^1_\BH/(2\ell)]$.
We thus get the familiar expression $E_b \approx -M^1_\BH M^2_\BH/\ell$.
If one of the objects is a NS, all that changes, is the individual energy $E_\NS$
associated with the star.  It is now obtained from the integration of the
NS energy density over the initial slice.  Including the contribution of
the BH to the volume element, the result
is $E_\NS \approx M_\NS(1+M_\BH/\ell) - E^\Self_\NS$,
where $E^\Self_\NS$ denotes the NS self-binding energy.  Therefore,
we do not recover the Newtonian expression.  The reason for this lies
in the way in which the NS has been introduced in the superposition and
the fact that the definition for its individual energy, $E_\NS$, does
not include the gravitational influence of the BH. It turns out that this way of
superposing the NS is closer to the case of a test body.  Indeed, if we look at the
binding energy by considering the NS as a test body, we find that $E_b = M_\NS(1-E_\NS/M_\NS)$,
i.\,e.\, the usual definition for test masses, and we obtain the correct Newtonian
expression.

We could modify our description by performing an additional
coordinate change to the NS model, prior to the superposition.
This would result in extra parameters, which may
be fixed so that the deformations of the NS produced by this coordinate
change will account for the effects caused by
the presence of the BH.  While a detailed study of such modifications is
beyond the scope of this paper, it will be interesting to study their
effect in future work.

\subsection{On the Equations of Motion for the NS}\label{nsmotion}

In a completely general setup the equations of motion for the matter
fields associated with
the NS just follow from the local conservation of
energy and momentum, i.e. $\nabla_\mu T^{\mu\nu}=0\,,$ which leads to
the equations of magneto-hydrodynamics.  In this paper, however, we
are advocating an approximate approach that avoids solving hydrodynamical
equations by reducing the number of physical degrees of freedom to a
finite number, the {\em hydro without hydro} approach.  More precisely,
the idea is, that we can approximate the matter description by a simplified
model that involves a finite number of parameters, $\{\lambda^I\}$, so
that the associated energy-momentum distribution is determined in terms
only of these parameters and the trajectory of a certain {\em center of
mass}, $z^\mu$, that is, $T^{\mu\nu}=T^{\mu\nu}[z^\rho\,;\lambda^I]\,.$
Then, in order to update the matter sources that enter in our BSSN
evolution equations we merely  need to evolve the parameters $\{\lambda^I\}$
(referred to in this work as the {\em internal} motion) and the
components of the trajectory $z^\mu$ (the {\em external} motion).
In practice, evolution of these parameters is obtained from ordinary
differential equations.
This is a direct consequence of going from an infinite number degrees of
freedom, as described by partial differential equations, to a finite number,
described by ordinary differential equations.

For the case of extreme-mass-ratio binaries, the structure of the
small object may be neglected as a first approximation,
so that we do not need to consider the interior motion. The external motion,
in turn, is given by the
solution of the geodesic equations in the numerically constructed spacetime.
In order to obtain a more realistic description of
BH-NS binaries with similar masses, it is important to take into account
the internal motion or, in other words, we need to allow for deformations
of the NS associated with tidal forces arising from the presence of the BH.
To that end, an interesting approach is to consider the
so-called affine stellar model introduced by Lattimer and
Schramm~\cite{Lattimer:1976ls} and further developed by Carter and
Luminet~\cite{Carter:1983cl,Carter:1985cl} and generalized for curved
spacetimes in~\cite{Luminet:1985lm}.  As it stands now,
however, the affine model is applicable only to slowly varying general
relativistic spacetimes~\cite{Mashhoon:1975ma,Mashhoon:1977ma,Luminet:1985lm},
where a stationary background can be identified.  This is an
assumption that cannot be adopted in our framework, since we
are interested in situations that involve strong and dynamical
gravitational fields.   Therefore, the affine model should be
generalized appropriately for dynamical spacetimes.
A particularly interesting way of doing this is based on Dixon's work
\cite{Dixon:1970wg,Dixon:1970ww,Dixon:1974wg,Dixon:1977wg} (see
also~\cite{Mashhoon:1977ma}).  The idea is to use the multipole moments associated
with the energy-momentum tensor to construct equations of motion for
the {\em center of mass}, i.\,e.\,$z^\mu$, and for the extra parameters that
describe the NS internal motion.  In any case, the affine model has been
used successfully to study the tidal interaction and disruption of stars
by supermassive BHs~\cite{Carter:1983cl,Luminet:1985lm,Diener:1995dk},
isolated rotating stars~\cite{Lai:1994lr}, stars in binary
systems~\cite{Lai:1995ls} and, more recently, gravitational signals
arising from tidal interactions~\cite{Casalvieri:2005hp}.  Another
semi-analytic approach is the ellipsoidal energy variational
or Roche-Riemann model~\cite{Lai:1993lr} and its relativistic
generalization~\cite{Wiggins:2000wl} which is formally equivalent to
the hydrostatic limit of the affine model.

In this paper, our goal is to investigate the stability of the numerical
relativistic evolutions under the presence of moving matter distributions.
To that end, we adopt the simplest possible evolution for the NS, namely
a fixed trajectory around the BH.

\section{Computational Framework and Numerical Computations}
The numerical results presented in this work have been obtained with
the 3D numerical relativity {\sc Maya} code which has
been described in detail in \cite{Shoemaker2003,Sperhake2004}. The
code uses the {\sc Cactus} computational toolkit~\cite{cactus} for
parallelization, data input/output and horizon finding. Inside the
{\sc Cactus} environment mesh refinement is provided by the {\sc
Carpet} package~\cite{carpet}. With regard to earlier versions
of the {\sc Maya} code used for BH simulations, the evolution
of extreme-mass-ratio binaries in the framework of the
geodesic approximation method has made necessary some additions and
modifications to the code.  Foremost, these are the implementation of
dynamic (moving) mesh-refinement and the addition of the matter
source terms on the right hand sides of the Einstein field equations. Both
will be described in more detail as we discuss the numerical simulations
performed in this work.

\subsection{Description of the Numerical Implementation}
\label{numericalframework}
We first address the calculation of the TOV profiles.
The TOV equations~(\ref{toveq1})-(\ref{toveq3}) constitute  a
boundary value problem for the variables $A(r)$, $m(r)$ and
$\rho(r)$.  This system of equations is closed by an equation of
state $p=p(\rho)$, chosen to be of polytropic type in this work. The
boundary conditions are given by $m=0$ at the center as well as $p=0$
and $A=\sqrt{1-2M^{}_\NS/R^{}_\NS}$ at the stellar surface.  We solve
the resulting system of ordinary differential equations using a standard
relaxation scheme~\cite{Press:92nr}.  In practice, we calculate the TOV
profiles during the initialization phase of the code and store them
as functions of the radial variable $r$. The values required on the
three-dimensional Cartesian grid are then obtained from third order
polynomial interpolation.

The discretization of the BSSN evolution
equations~(\ref{dtgdt}-\ref{dtGdt}) has been implemented using centered
second order stencils for spatial derivatives, except for the advection
terms of type $\beta^k \partial_k$, for which second order accurate
upwinding operators have been used. The integration in time is done
using an iterative Crank-Nicholson scheme. As in simulations of vacuum
spacetimes we have treated the spacetime singularities associated
with the BHs by excising an area of finite size inside the apparent
horizons (see~\cite{Shoemaker2003,Sperhake2004} for the details of
the implementation of this technique).

\subsection{Analysis of the Initial Data}\label{analysisinitialdata}

In order to study the properties of the initial data proposed in
this paper we have considered six different TOV models for the NS.
The physical features of these models (the polytropic coefficient $k$,
the polytropic index $n$, stellar radius $R^{}_\NS$ and mass $M^{}_\NS$,
compactness ratio $M^{}_\NS/R^{}_\NS$, and central density $\rho^{}_{\rm
c}$) are shown in Table~\ref{tab:stellar_models}.  Notice, that the
compactness ratios are small and hence some of the models would describe
stars much less compact than astrophysical NSs.

\begin{table}
  \begin{center}
  \caption{\label{tab:stellar_models}
           Parameters and resulting mass and density
           for the six stellar models discussed in this work.}
  \begin{ruledtabular}
  \begin{tabular}{c|ccc|ccc}
   ~~ Model~~ & $k$ & ~$n$~ & ~$R^{}_\NS\;[M_\BH]$~~ & ~$M_\NS\;[M_\BH]$~ &
    $M_\NS/R_\NS$ & $\rho_{\rm c}\,[M^{-2}_\BH]$~~ \\[1mm]
  \hline\\[-3mm]
    1 &  $0.831$     & $1$ &  $1$  & $10^{-1}$ & $10^{-1}$ & $7.9\,10^{-2}$ \\
    2 &  $0.652$     & $1$ &  $1$  & $10^{-2}$ & $10^{-2}$ & $7.9\,10^{-3}$ \\
    3 &  $0.638$     & $1$ &  $1$  & $10^{-3}$ & $10^{-3}$ & $7.9\,10^{-4}$ \\
    4 &  $0.00831$   & $1$ & $0.1$ & $10^{-2}$ & $10^{-1}$ & $7.9$ \\
    5 &  $0.00652$   & $1$ & $0.1$ & $10^{-3}$ & $10^{-2}$ & $7.9\,10^{-1}$ \\
    6 &  $0.00638$   & $1$ & $0.1$ & $10^{-4}$ & $10^{-3}$ & $7.9\,10^{-2}$ \\
  \end{tabular}
  \end{ruledtabular}
  \end{center}
\end{table}

From the values in Table~\ref{tab:stellar_models} we find that the central
density does not only increase with the compactness of the stellar model,
but also when we make the star smaller, keeping constant the compactness
ratio (see, for instance, models 1 and 4). This is an expected result:
if we reduce the mass $M^{}_\NS$ while keeping the compactness ratio
$M_\NS/R_\NS$ constant, the density behaves like
\begin{align}
\rho \sim \frac{M^{}_\NS}{R^3_\NS} = \frac{1}{M^2_\NS} \frac{M^3_\NS}{R^3_\NS}
= \frac{\mathrm{const}}{M^2_\NS}\,.
\end{align}
That is, the density scales as the inverse of the square of the mass.
The values in Table~\ref{tab:stellar_models} confirm this relation.
From the computational point of view, this feature represents a
considerable challenge.  For more extreme mass ratios (smaller
$M^{}_\NS$) we do not only need to decrease the grid spacing linearly
with the size of the star.  Further increase in resolution is necessary
to adequately resolve the steep gradients resulting from the increasingly
larger matter densities and curvature encountered for smaller values of
$M^{}_\NS$.  Even with the availability of mesh refinement, as discussed
below, the computational demands for simulating low mass compact sources
quickly become prohibitively costly.  It is for this reason that we also
consider the less compact models listed in Table~\ref{tab:stellar_models}.
We emphasize, however, that all models under consideration here are
orders of magnitude more compact than those used in similar studies
in the literature~\cite{Bishop:2003bs}.  Below we will discuss the numerical
demands arising from high compactness more quantitatively, when we discuss
the constraint violations of the superposed data.  First, we test the
numerical implementation of the boosted TOV solution by studying the
convergence of the constraints in the absence of a BH.

The initial data constructed according to the procedure described above
results in exact solutions of Einstein's equations in various limits.
In particular, by setting the mass of the BH to zero we recover
the solution of a boosted TOV star in KS-like coordinates.  We emphasize,
that, in the absence of the BH, there is no fundamental length scale
$M^{}_\BH$, so that the numerical values obtained for the NS no
longer have the unambiguous meaning as before.  A key motivation for this
study, however, is to probe the suitability of the numerical framework
used for the simulations.  For this purpose we construct the numerical
grid using the same numerical values as if a BH of mass $M^{}_\BH$ were present.
In this work we limit our discussion to non-spinning BHs and
stellar models.  In consequence, the configurations under study are
inherently symmetric about the orbital plane.   It is sufficient,  for
this reason, to evolve data only in the bitant $z>0$ and impose symmetry
conditions in the orbital $x,y$ plane.

In order to achieve maximum resolution at moderate computational
cost, we use Fixed Mesh Refinement for all simulations in this work.
Inside the MAYA code, Mesh Refinement is implemented via the CARPET
package~\cite{carpet,Schnetter:2004sh}.   In contrast to the static fixed
mesh refinement used for the head-on collisions in~\cite{Sperhake:2005uf}
the type of scenario investigated here will eventually require dynamic
mesh refinement to accommodate the orbital motion of the stellar object
around the BH. We return to this issue later, when we will
discuss time evolutions.  For the initial data under consideration here,
we restrict our attention to the initial setup of the refinement levels.
The initial position of the star in this case is $x=8.64$, $y=z=0$ and
it has an initial boost velocity given by $v_x=v_z=0\,,$ $v_y=0.34$.
For the stellar models 1-3 of Table~\ref{tab:stellar_models} we use a
set of 5 nested boxes around the BH plus a set of two nested
refinement components centered around the NS.  For models 4-6
we add three further refinement levels centered on the NS.
The exact specifications of the refinement levels are given in
Table~\ref{tab:reflevels}.

\begin{table}
  \begin{center}
  \caption{\label{tab:reflevels}
           Structure of the refinement levels as used in the study of
           the initial data. The two values for the grid spacing
           correspond to the coarse and fine resolution in the
           convergence analysis. The boundaries in the $z$-direction
           are $z_{\rm min}=0$ and $z_{\rm max}=y_{\rm max}$ on all
           levels and have not been listed explicitly. Levels 7-9 are
           only used for stellar models 4-6.}
    \begin{ruledtabular}
    \begin{tabular}{c|cc|cccc|cccc}
    Refinement &&& \multicolumn{4}{c|}{Grids centered around the BH} &
                 \multicolumn{4}{c}{Grids centered around the NS} \\
    Level      & $h_1\;[M_\BH]$ & $h_2\;[M_\BH]$  &  $x_{\rm min}\;[M_\BH]$ & $x_{\rm max}\;[M_\BH]$
                                &  $y_{\rm min}\;[M_\BH]$ & $y_{\rm max}\;[M_\BH]$~~
                                &  $x_{\rm min}\;[M_\BH]$ & $x_{\rm max}\;[M_\BH]$
                                &  $y_{\rm min}\;[M_\BH]$ & $y_{\rm max}\;[M_\BH]$~~\\[1mm]
    \hline\\[-3mm]
    1 & 2.88 & 1.92 & -103.68 & 103.68 & -103.68 & 103.68 &        &         &          &          \\
    2 & 1.44 & 0.96 &  -51.84 &  51.84 &  -51.84 &  51.84 &        &         &          &          \\
    3 & 0.72 & 0.48 &  -25.92 &  25.92 &  -25.92 &  25.92 &        &         &          &          \\
    4 & 0.36 & 0.24 &  -12.96 &  12.96 &  -12.96 &  12.96 &        &         &          &          \\
    5 & 0.18 & 0.12 &   -4.32 &   4.32 &   -4.32 &   4.32 &  6.48  &  10.80  &   -2.16  &   2.16   \\
    6 & 0.09 & 0.06 &         &        &         &        &  7.56  &   9.72  &   -1.08  &   1.08   \\
    7 & 0.045& 0.03 &         &        &         &        &  8.1   &   9.18  &   -0.54  &   0.54   \\
    8 & 0.0225& 0.015&        &        &         &        &  8.37  &   8.91  &    0.27  &   0.27   \\
    9 & 0.01125& 0.0075&      &        &         &        &  8.505 &   8.775 &    0.135 &   0.135  \\
  \end{tabular}
  \end{ruledtabular}
  \end{center}
\end{table}

For the convergence analysis we have calculated the data for two
different grid spacings $h_1$ and $h_2=h_1/1.5$ as listed in the table. Because
an isolated boosted TOV star is a solution of the Einstein equations we
expect the Hamiltonian and momentum constraints, given by
equations~(\ref{Hconstraint},\ref{Pconstraint}),
to converge to zero at second order. Numerically this implies that
the constraint violations obtained for the high resolution calculation
should be $1.5^2$ smaller than those obtained at coarse resolution.

\begin{figure}
  \begin{center}
  \includegraphics[height=400pt,angle=-90]{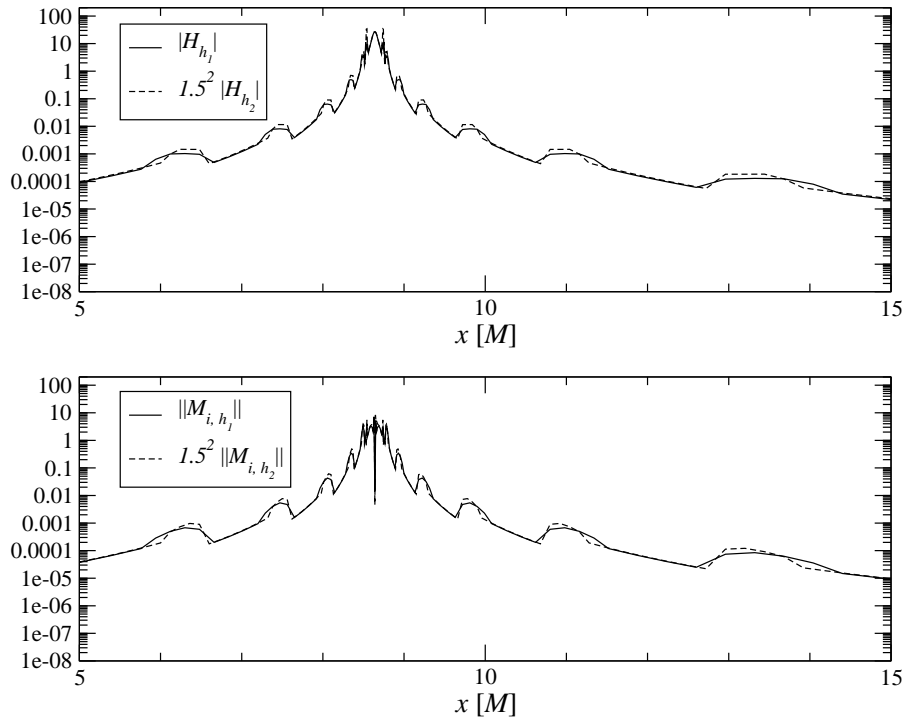}
  \caption{Constraint violation of the Hamiltonian and the
           norm of the momentum constraint along the $x$-axis
           for stellar model $4$ in the absence
           of a BH. The constraints obtained at high resolution
           have been amplified by $1.5^2$.}
  \label{fig:close_r0.1_m0.01_hamx_momxx}
  \end{center}
\end{figure}

Figure~\ref{fig:close_r0.1_m0.01_hamx_momxx} demonstrates second order
convergence for the Hamiltonian and the
norm $||\mathcal{M}||= \sqrt{\sum_i{\mathcal{M}_i^2}}$
of the momentum constraints
along the $x$-axis, obtained for the most compact model of Table
\ref{tab:stellar_models}, model 4.  We find similar results
for all constraint functions on arbitrary axes. A similar
analysis at coarser resolution ($1.5\,h_1$ instead of $h_2$), however,
gave less satisfactory results, indicating that the resolutions used here
are necessary for model 4. For the other, less extreme, models we find
correspondingly less strict requirements on the resolution to achieve
second order convergence.

We now turn our attention to the binary data obtained for non-vanishing
mass of the BH. In the continuum limit the resulting data are
a solution of the Einstein equations only as the separation of the two
objects becomes infinite. We therefore do not expect the constraint violations
to converge to zero. On the other hand, the constraints will never be
satisfied exactly in a numerical simulation because of the finite
numerical accuracy. In order to assess the significance of the constraint
violations inherent to the superposed data relative to the limitations
in numerical accuracy we follow the same approach as in~\cite{Sperhake2004}
and study the constraint violations at different resolutions. If the
constraint violations decrease at about second order for higher resolutions
we conclude the inherent constraint violations (due to the fact that we 
are not solving them) to be insignificant relative to the purely numerical error.

\begin{figure*}[t]
  \begin{center}
   \includegraphics[height=250pt,angle=-90]{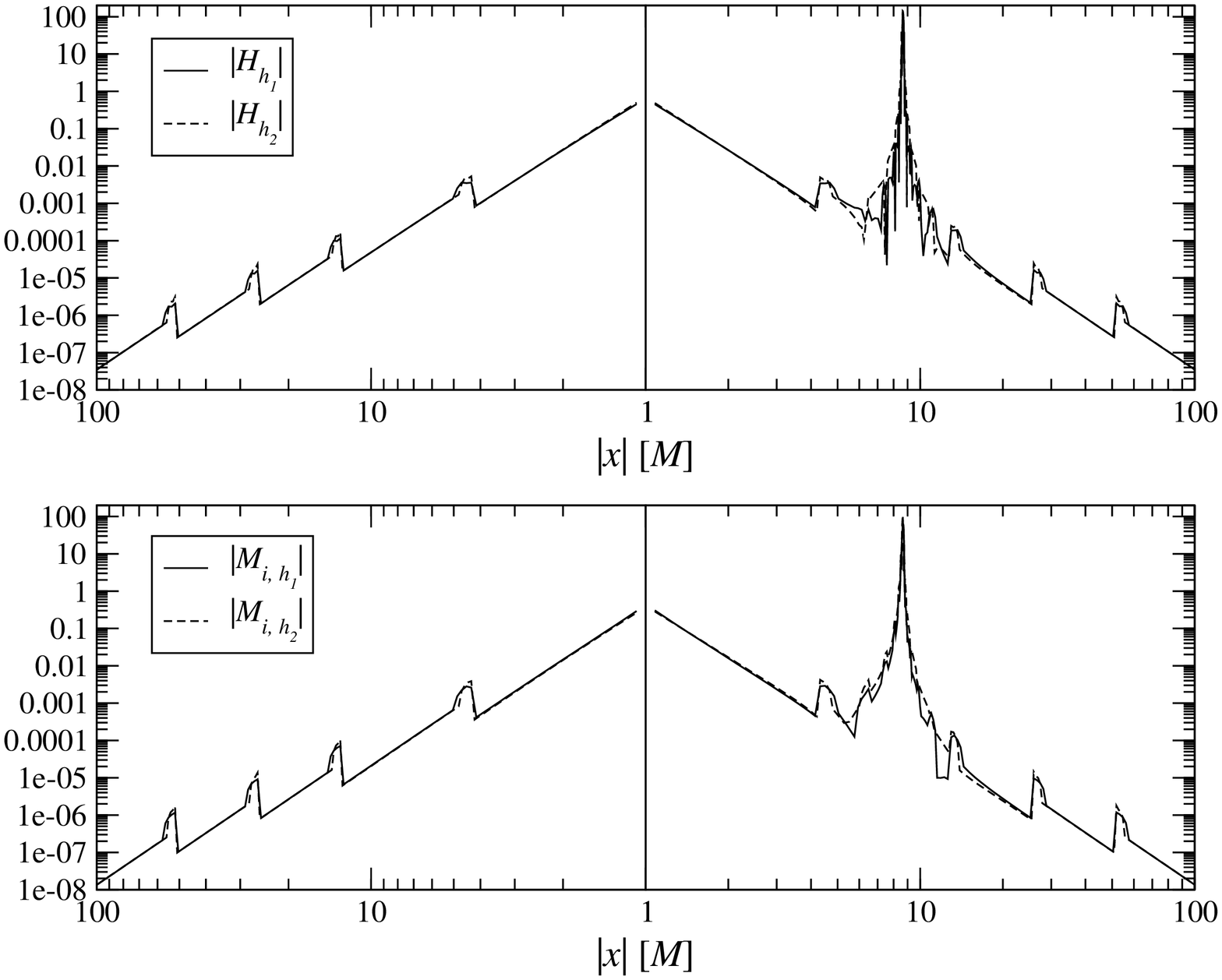}
   \includegraphics[height=250pt,angle=-90]{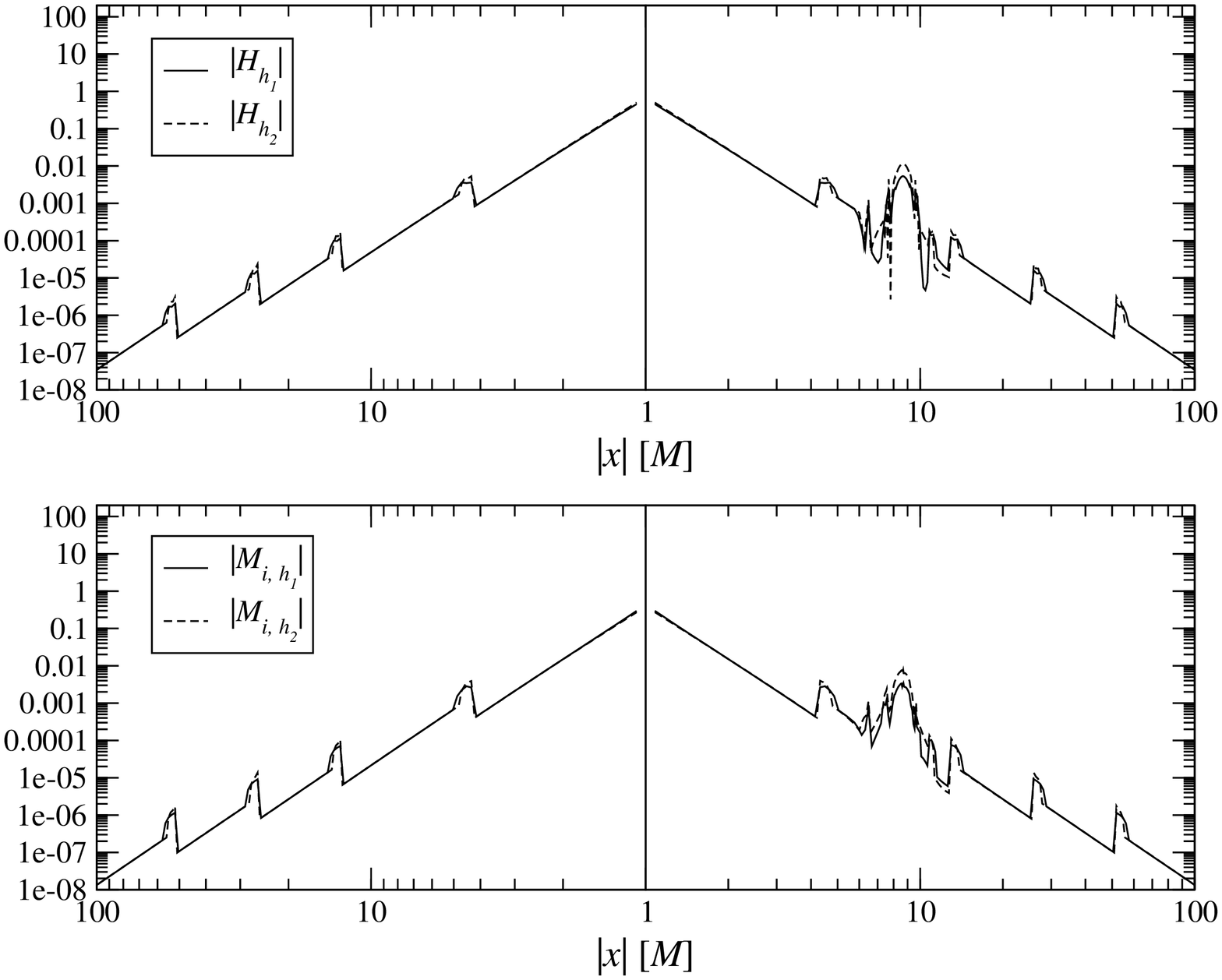}
  \caption{Constraint violation of the Hamiltonian
           and momentum constraints along the $x$-axis
           for stellar model $4$ (left panels) and $3$ (right panels).
           The constraints obtained at high resolution
           have been amplified by $1.5^2$.}
  \label{fig:constraints_bh}
  \end{center}
\end{figure*}

For this purpose we show in Figure~\ref{fig:constraints_bh} the convergence
of the Hamiltonian constraint and the norm $||\mathcal{M}||$
of the momentum constraints along the $x$-axis.
For clarity we have used a logarithmic scale for the distance from the
origin in both directions. As in the
case of an isolated star, we find that the constraints converge away
at second order for the resolutions used here. The only exception is the
region of the star, where the non-linear interaction between the
gravitational fields of the two objects is strongest and we expect
the constraint violations inherent to this data construction to
be largest. It is these constraint violations which represent the
approximative nature of our approach to study this type of binary
configurations.

Figure~\ref{fig:constraints_bh} also illustrates the high computational
demands for simulations of more compact objects. The two models (3 and 4
of Table \ref{tab:stellar_models} used in the figure represent the
least and most compact stars. Clearly the constraint violations found
for model $4$ in the upper two panels are very large compared to the
average violations near the BH. This is an artifact of the high
densities and curvature encountered near the small star in combination
with resolutions currently available with our computers.
In contrast the constraint violations found for model $3$ in the lower
panels is substantially smaller than those observed near the BH.
We emphasize that this issue is separate from the convergence properties
mentioned above. In fact, the convergence properties found for model 4
are closer to second order than those found for model $3$. We attribute
this to the fact that the constraint violations near the very compact
object $4$ are dominated by the discretization errors arising from the
insufficient resolution near the stellar center. For these reasons
we study in the next subsection evolutions of the less compact model $3$ only,
for which we are confident to achieve sufficient resolution, even inside
the star.

\subsection{Time evolutions}\label{timeevolutions}
In this section we present time evolutions of the initial data presented above.
Following our previous discussions, we focus on the implementation of the 
different elements of this type of simulations in the framework of moving 
refinement components with emphasis on the stability of the evolutions.

To this end, we have evolved the superposed data of a BH of mass
$M_\BH=1$ and stellar model $3$ as listed in Table \ref{tab:stellar_models}.
Because of memory restrictions of the available computational resources we
have evolved this binary system using a distance $d=7\,M$ and a grid setup
slightly different from that of Table \ref{tab:reflevels}. The
grid specifications used in the evolution are listed in Table
\ref{tab:reflevels_evol}. Of the refinement levels listed there,
number $1$ to $5$ remain fixed throughout the evolution. Level $6$, however,
is required to follow the motion of the stellar object and thus needs to
move. For this purpose we use the {\em regridding} operation
implemented inside {\sc Carpet}. This operation allows for interpolation
of function values from coarser onto finer refinement levels. This
operation is regularly performed in mesh refinement simulations of
Berger-Oliger type to provide boundary conditions for the inner
levels, the so-called {\em prolongation}. In case of a moving high
resolution grid component, the same operation is used to fill the
new points on that component with valid data from the coarser levels.
Inside the {\sc Maya} code we use this feature by adjusting the
specifications, i.e.\,$x_{\rm min}$, $x_{\rm max}$, $y_{\rm min}$
and $y_{\rm max}$, of refinement level $6$ in the course of the evolution.

\begin{table}
  \begin{center}
  \caption{\label{tab:reflevels_evol}
           Structure of the refinement levels as used for the
           time evolution.}
  \begin{ruledtabular}
  \begin{tabular}{c|c|cccc}
    Refinement Level      & $h\;[M_\BH]$ &  $x_{\rm min}\;[M_\BH]$ & $x_{\rm max}\;[M_\BH]$
                          &  $y_{\rm min}\;[M_\BH]$ & $y_{\rm max}\;[M_\BH]$ \\
    \hline\\[-3mm]
    1 & $4.00$  & $-144.00$ & $144.00$ & $-144.00$ & $144.00$  \\
    2 & $2.00$  & $ -72.00$ & $ 72.00$ & $ -72.00$ & $ 72.00$  \\
    3 & $1.00$  & $ -36.00$ & $ 36.00$ & $ -36.00$ & $ 36.00$  \\
    4 & $0.50$  & $ -18.00$ & $ 18.00$ & $ -18.00$ & $ 18.00$  \\
    5 & $0.25$  & $ -12.00$ & $ 12.00$ & $ -12.00$ & $ 12.00$  \\
    6 & $0.125$ &  $5.00$   & $9.00$   & $-2.00$   & $2.00$    \\
  \end{tabular}
  \end{ruledtabular}
  \end{center}
\end{table}

We have implemented two alternative ways of controlling these values.
The first uses the tracking of the BH singularity, as provided,
for example, by an apparent horizon finder, and moves the center of the
refinement component by a corresponding amount. Because this method is
inherently restricted to BHs, we also allow for the center of the
component to follow a user specified trajectory. This second method is
realized for the case of the stellar object discussed in this work. The
\begin{figure}[t]
  \begin{center}
   \includegraphics[height=170pt,angle=-0]{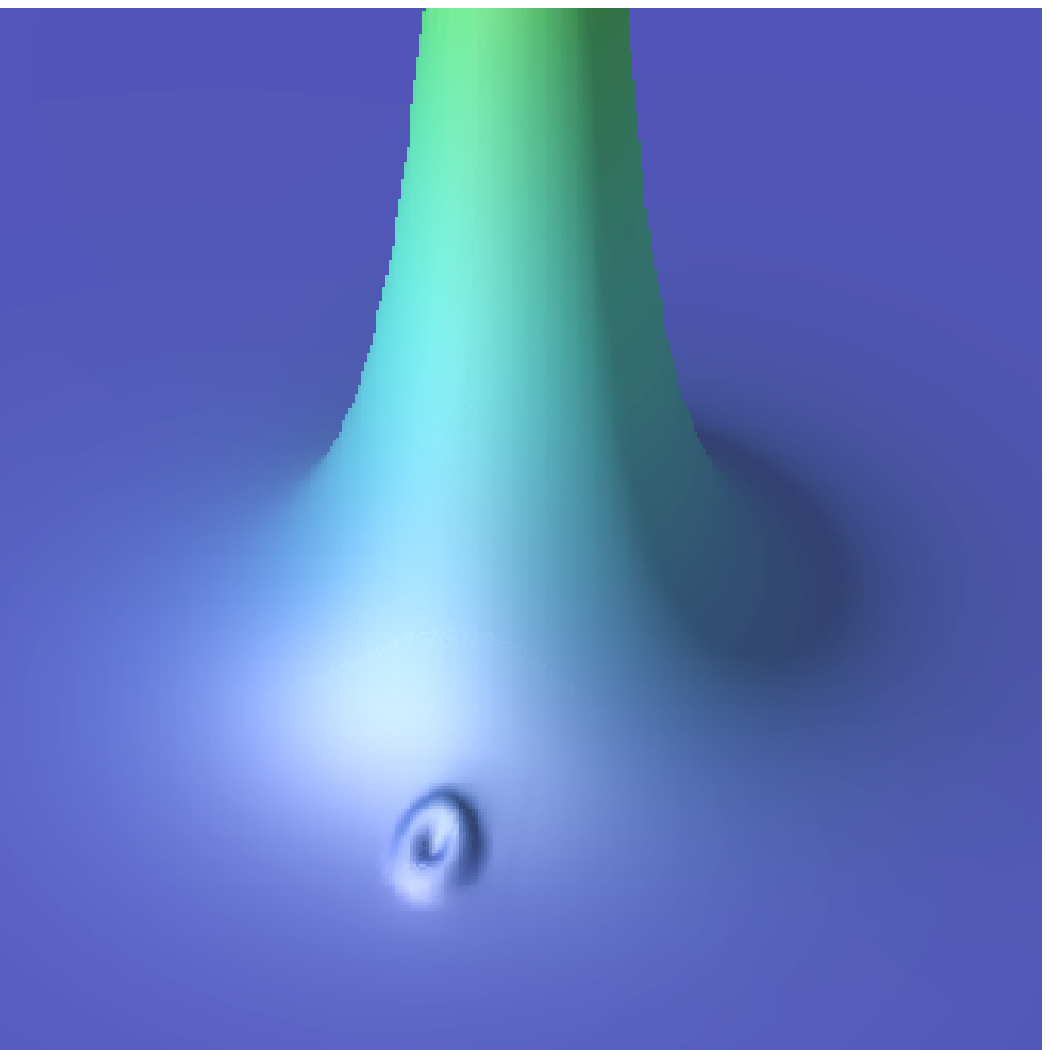}
   \includegraphics[height=170pt,angle=-0]{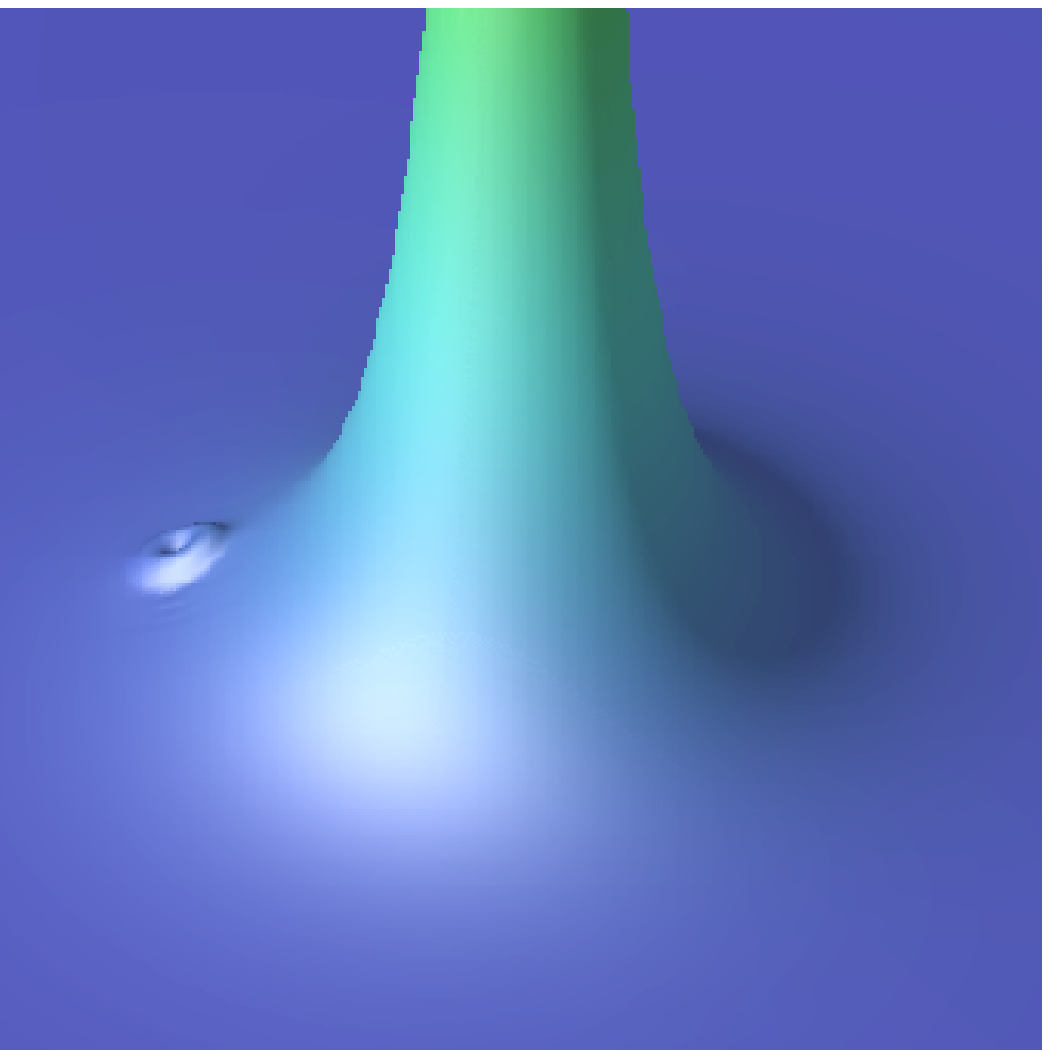}
   \includegraphics[height=170pt,angle=-0]{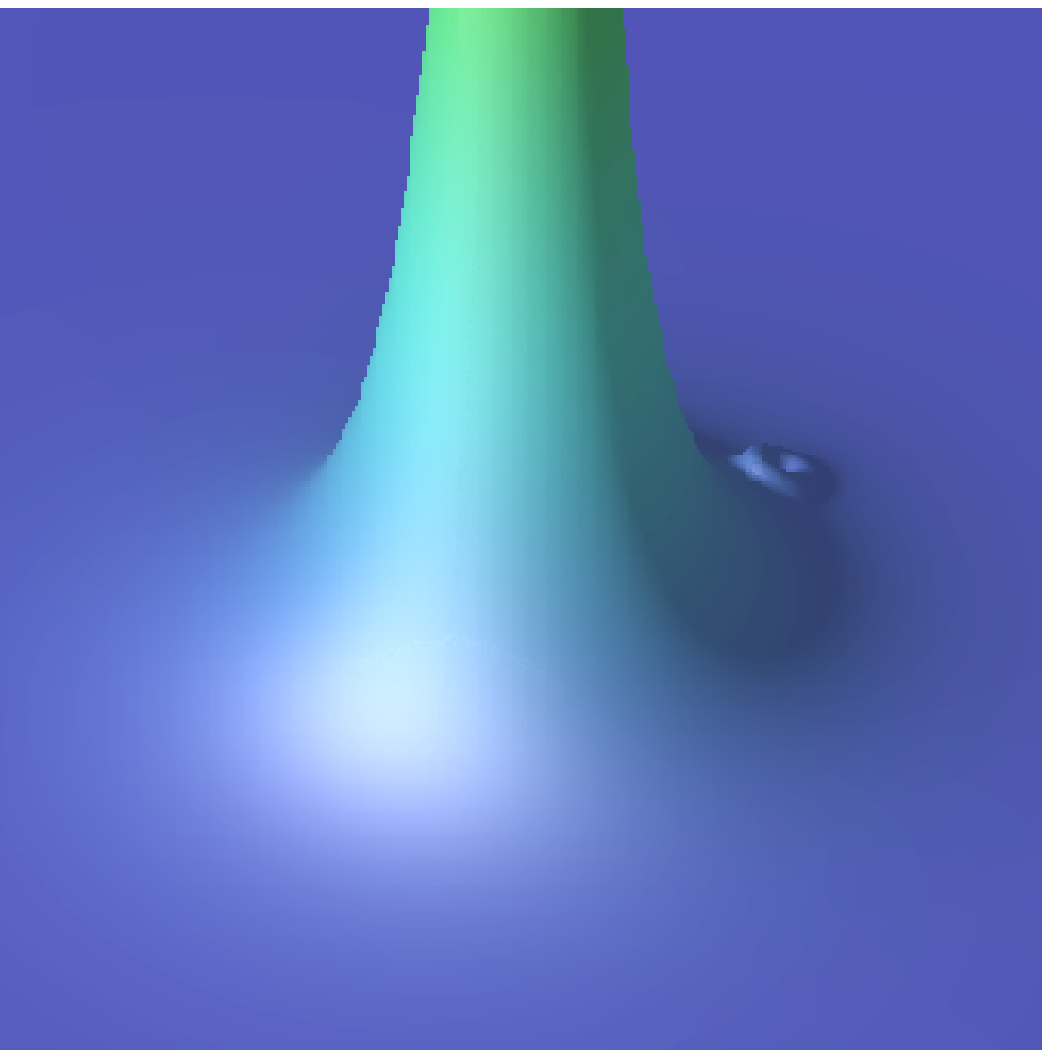}
   \includegraphics[height=170pt,angle=-0]{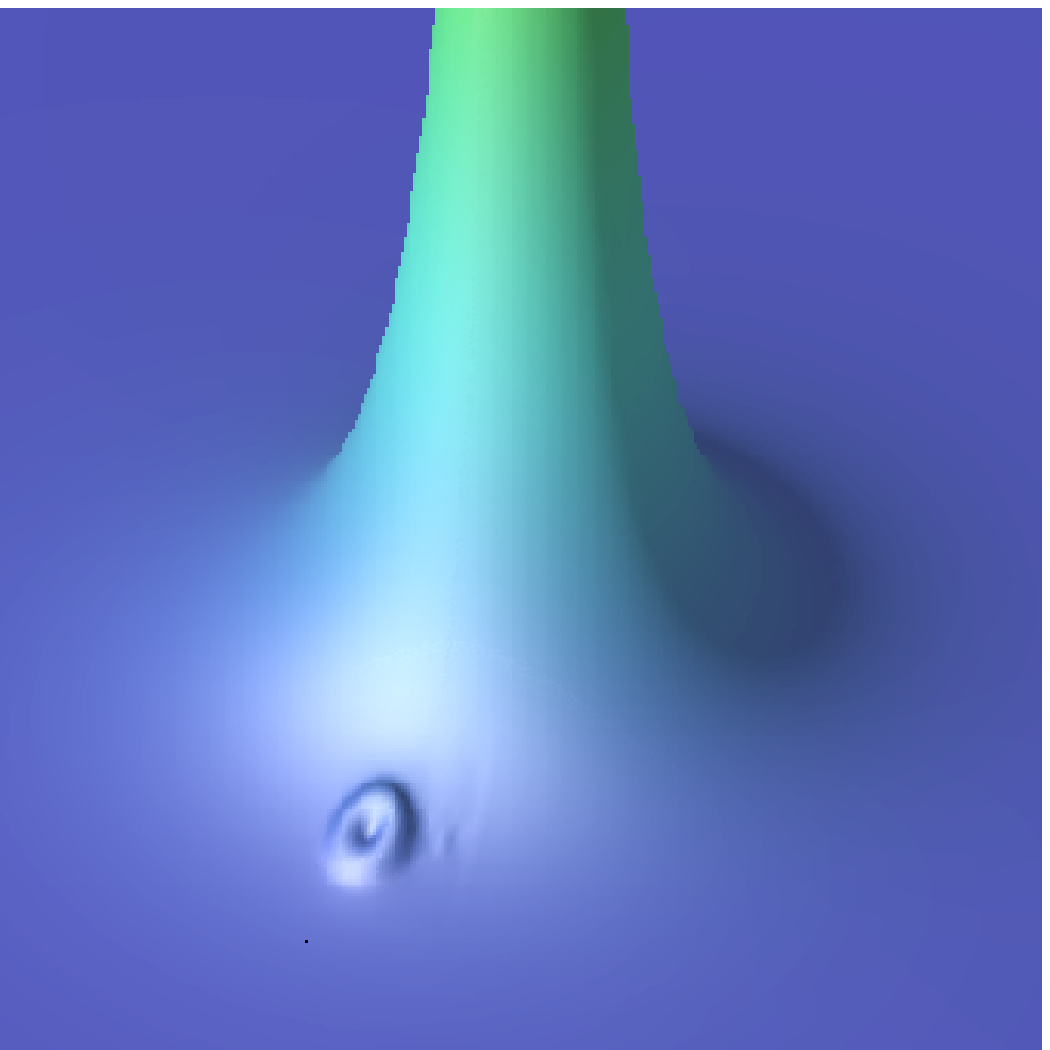}
  \caption{Snapshots of the time evolution of a NS model type
           model 3 orbiting a massive BH with orbital frequency
           $\omega=0.054\,M_\BH^{-1}$.}
  \label{fig: evolution}
  \end{center}
\end{figure}
trajectory is the same as that used for the motion of the
stellar object itself. In Fig.\,\ref{fig: evolution} we show four snapshots
of the ensuing evolution. The figure shows the trace of the extrinsic
curvature obtained at times $t=0$, $30\,M$, $70\,M$ and $120\,M$. While the
BH is represented by the large central throat, the stellar
object manifests itself in the form of the small perturbation
initially visible to the right of the BH (upper left panel).

The simulations we have carried out last at least for 
an evolution time of $167\,M$, corresponding to about one and a half
orbits, without showing signs of instability.
Furthermore, no elaborate fine-tuning of the parameters was necessary to
achieve evolutions.   In conclusion, simulations on orbital timescales appear 
to be achievable rather straightforwardly using the approach discussed in this 
work.

\section{Conclusions \label{sec:conclusions}}

In this paper we have introduced a framework to study BH-NS binaries
using a fully non-linear evolution of the spacetime geometry and
an approximate evolution of the NS matter sources based on the
reduction of the degrees of freedom of the NS, so that hydrodynamical
calculations are avoided. The remaining degrees of freedom, including
the motion of the NS are governed by ODEs.   We expect this framework
to be useful to investigate certain dynamical regimes of the BH-NS
binary system relevant for the point of view of gravitational-wave
astronomy.

The construction of the initial data is performed by generalizing the
superposition technique of \cite{Matzner:1998pt}, as applied to
BH binaries in KS form, to BH-NS binaries in KS-like coordinates.
In order to make this approach functional in
numerical relativity using three-dimensional Cartesian coordinates,
we found it necessary to modify the standard coordinate transformation
to KS coordinates inside the NS.
In particular, we used a coordinate transformation that avoids the
appearance of a coordinate singularity at the stellar centre,
while preserving the
light-like geodesic structure of the coordinates outside the star.

We have studied the properties of the initial data thus obtained for
a set of stellar models with different values of the compactness and mass
ratios.  An adequate description of extreme-mass-ratio binaries
involving compact stars is only
made possible by the use of advanced mesh-refinement techniques. We have
demonstrated second-order convergence of the Hamiltonian and momentum
constraints in the case of an isolated NS (corresponding to an
infinite BH-NS separation). For finite separations we still observe
second-order convergence of the constraints over large parts of the
computational domain. We thus conclude that these initial data sets
are as suitable for the description of BH-NS binaries as the pure BH-BH analog
previously discussed in the literature.

Finally, we have investigated the feasibility of evolving such
scenarios in time using modern methods of numerical relativity. 
In the numerical simulations
considered in this work for testing purposes, we have restricted ourselves
to the simple case in which the NS follows a prescribed trajectory.
Foremost, our interest was focused on the stability of the resulting simulations
and the demands on computational resources. We have found that for
sufficiently extreme mass ratios, long-term stable simulations on
orbital time scales can be achieved straightforwardly using standard
evolution schemes for solving the Einstein field equations.
As the equal-mass limit is approached, the stability properties
deteriorate because
of the strong dynamics in the vicinity of the BH.
In this limit, our approach no
longer provides an adequate description of the matter sources, however.
The scenarios of immediate interest for our approximation technique are
thus within the range of capability of our evolution code.

With regard
to the compactness of the NS we find that the combination of
extreme mass ratios with high compactness of the NS result
in very steep gradients. These, in turn, demand a very high resolution
and a large number of refinement levels. Currently we do not have
the computational resources available to provide such resolution in long-term
evolutions. We have therefore studied the time evolutions of less
compact objects, but still orders of magnitude more compact than what
has hitherto been considered using this type of approach. Such
studies are well within the scope of current resources and can
be refined in the future, by adding radiation-reaction effects
to the dynamics and/or internal degrees of freedom of the NS to study
different aspects of these systems in the context of gravitational-wave
astronomy.

{\bf Acknowledgements:}
The authors acknowledge the support of the Center for Gravitational Wave Physics
funded by the National Science Foundation under Cooperative Agreement
PHY-0114375, and support from NSF grant PHY-0244788 to Penn State University.
The authors wish to thank Erik Schnetter for help with the implementation
of moving refinement boxes. U.S. acknowledges support from the
SFB/TR 7 ``Gravitational Wave Astronomy'' of the German Science Foundation
and the hospitality of the Albert Einstein Institute (Potsdam) during a visit.



\bibliographystyle{apsrev}


\end{document}